\documentclass[12pt]{iopart}
\usepackage{graphicx,color}
\usepackage{epsfig}
\usepackage[normalem]{ulem}
\usepackage{iopams} 
\newcommand{\vct}[1]{{\bf #1}}
\newcommand{\fig}[1]{figure~\ref{Fig:#1}}
\newcommand{\tab}[1]{table~\ref{tab:#1}}

\begin{document}

\title{Kinetic energies of liquid and solid phases of $^4$He}

\author{E.~J.~Rugeles,$^1$ Sebastian~Ujevic$^{2}$ and S.~A.~Vitiello$^{1}$}
\address{$^1$ Instituto de F\'{\i}sica Gleb Wataghin, Universidade Estadual de Campinas - UNICAMP 13083-859 Campinas, SP, Brazil}
\address{$^{2}$ Departamento de Ci\^encias Exatas - EEIMVR, Universidade Federal Fluminense 27255-125 Volta Redonda, RJ, Brazil}

\begin{abstract}
Kinetic energies of a system of $^4$He are investigated at zero temperature. The multi-weight extension to the diffusion Monte Carlo method is used to implement the Feynman-Hellmann theorem in an effective way. This method allows the quantities of interest to be computed with excellent accuracy. In order to study the importance of symmetry in the kinetic energy calculations, we have considered for the solid phase two guiding wave functions: the Nosanov-Jastrow without boson symmetry and the symmetric Nosanov-Jastrow with boson symmetry. In general very good agreement is found with the experimental data at both the liquid and solid phases.
\end{abstract}
\maketitle

\section{Introduction}

In quantum systems single particle momentum distributions $n(\vct k)$ and their associated kinetic energies $E_k$ are of great interest for both experimental and theoretical physicists. These quantities can characterize the extent in which these systems deviates from classical physics. The strong quantum effects present in the condensed phases of $^4$He atoms offer an example where  the $n(\vct k)$ can not be described by the Maxwell-Boltzmann distribution  typical of classical systems.

The most widely studied property of solid helium using neutrons is the kinetic energy \cite{gly13}. To date,  only experimental techniques associated with neutrons allow for direct measurements of the kinetic energy and the momentum distribution of this system. In the liquid phase these quantities have also been the aim of many theoretical studies.

Estimations of the ``exact" kinetic energy at zero temperature can be made through the diffusion Monte Carlo (DMC) method. However its computation is not straightforward  as for the total energy. This is because the kinetic energy is a quantity that does not commute with the Hamiltonian of the system. A possible approach to  estimate the kinetic energy is through the Hellmann-Feynman theorem using the multi-weight diffusion Monte Carlo method \cite{vit11jcp,vit11jltp} that allows the involved derivatives to be performed in a reliable way. In this extension of the standard DMC, the potential energy can be estimated without the usual difficulties associated with numerical differences of independent estimates and  uncorrelated statistical uncertainties. In general, determinations of the kinetic energy might also be technically relevant in the context of projection Monte Carlo methods. This is because the total energy might have a bias proportional to the expectation value of the kinetic energy \cite{cer95}.

In this work we estimate the total and kinetic energies for the liquid and solid phase of a system formed by $^4$He atoms. In the solid phase, we have explored in what degree two guiding functions with different overlaps with the true ground state would affect the results for the kinetic energy of the system. We have experienced with both a guiding function of the Nosanov-Jastrow (NJ) form and the so-called symmetric Nosanov-Jastrow (SNJ) function \cite{caz13,caz09}. We want to verify how the introduction of a symmetric guiding function in the calculations would change the results obtained with a  function that do not has this property.

In the next section we describe the helium atoms system considered in this work. We also present the way in which the Feynman-Hellmann theorem is implemented in the multi-weight diffusion Monte Carlo (DMC) method to compute kinetic energies. Results are given and some final comments are made in the last sections.

\section{Method}

We want to compute the kinetic energy of a system formed from $^{4}$He atoms. With this aim we consider $N$ bodies using periodic boundary conditions described by the Hamiltonian

\begin{equation}
H=-\frac{\hbar^{2}}{2m}\sum_{i=1}^{N}\nabla_{i}^{2}+\sum_{i<j} V\left(r_{ij}\right),
\label{eq:Hamiltonian}
\end{equation}

\noindent where  $V\left(r_{ij}\right)$ is the interaction potential that depends on the inter-atomic distance $r_{ij}=\left|\vct{r}_{i}-\vct{r}_{j}\right|$. We have considered the \textsc{hfd-b3-fcci1} potential proposed by Aziz and coworkers \cite{azi95}

\begin{equation}
V\left(x=r/r_{m}\right)=\epsilon\left[A\mbox{e}^{-\left(\alpha x+\beta x^{2}\right)}-B\left(x\right)\sum_{j=0}^{2}\frac{C_{2j+6}}{x^{2j+6}}\right]\label{eq:Aziz-95}
\end{equation}

\noindent where $r_{m}$ is the position of the potential minimum. The values of $\epsilon$, $\alpha$, $\beta$ and $C_{2i+6}$ can be found in \cite{azi95}. The function $B\left(x\right)$ is given by

\begin{equation} B(x) =\left\{
\begin{array}{ll}
\e^{-(\frac{D}{2}-1)^2} & \mbox{if $x$ $\leq$ $D$},\\
                      1 & \mbox{otherwise,}
\end{array}
\right.
\label{eq:Dunping-function-Potential-Aziz-95}
\end{equation}

\noindent where $D$ is another parameter of the inter-atomic potential.

A standard method to determine the ground-state energy of this system is the DMC method \cite{mos82,rey82}. This is accomplished by the simulation in imaginary time of a corresponding classical diffusion process with a source $V(r)$. This process is guided by a trial wave function able to give a meaningful description of the ground state of the system. For efficiency, branching of configurations is introduced to avoid the burden of keeping those configurations that practically do not give any contribution to the results.

In the solid phase we have experienced with two different guiding functions. One of them is of the Nosanov-Jastrow form

\begin{eqnarray}
\Psi_{NJ}  &=\psi_J\left(\vct{R}\right)\Phi\left(\vct{R}\right),\\
\psi_J\left(\vct{R}\right) & =  \prod_{i<j}^{N}\mbox{e}^{-\frac{b}{2}\left(\frac{1}{r_{ij}}\right)^{5}},\\
\Phi\left(\vct{R}\right) & =  \prod_{i=1}^{N}\mbox{e}^{-\frac{c}{2}|\vct{r}_{i}-\vct{l}_{i}|^{2}},
\end{eqnarray}

\noindent where $\vct{R}=\left\{ \vct{r}_{1},\cdots,\vct{r}_{N}\right\} $, $\vct{l}_{i}$ are the lattice sites of the crystal, and finally $b$ and $c$ are variational parameters. A second guiding function was also considered in order to infer the importance of a symmetric guiding function in the calculations of the kinetic and potential energies. We choose to employ a symmetric Nosanov-Jastrow function given by

\begin{equation}
\Psi_{SNJ}(\vct R) = \psi_J(\vct R)\prod_j\sum_i \e^{-\frac{c}{2}|\vct r_i - \vct l_j|^2},
\end{equation}

\noindent already used in DMC calculations \cite{caz13,caz09}. It is important to note that this is an approximation to a true symmetric Nosanov-Jastrow wave function. In the liquid phase a Jastrow factor $\psi_J(\vct R)$ was used as a guiding function. 

Since the kinetic energy operator does not commute with the Hamiltonian, its estimation is not straightforward  in a DMC calculation. For this reason, we have implemented the Hellmann-Feynman theorem in the multi-weight extension of this method. The kinetic energy is then obtained from the estimated values of the potential and total energies.

The potential energy $E_{p}$ is computed through the substitution $V\rightarrow\lambda V$ in the Hamiltonian and  the total energy $E^{\lambda}$ derivative with respect to the parameter $\lambda$

\begin{equation}
E_{p}=
\frac{d}{d\lambda}\left\langle H{\left(\lambda\right)}\right\rangle \Big|_{\lambda=1}.
\label{eq:Helman-Feynman-Theorem}
\end{equation}

\noindent Finally the kinetic energy $E_k$ is computed through 

\begin{equation}
E_{k}=E^{\left(\lambda=1\right)}-E_{p}.
\end{equation}

In the implementation of the Hellman-Feynman theorem through the multi-weight extension to the DMC method \cite{uje03,vit11jcp}, three different weights are associated to a given configuration or walker $\vct{R}$, one for each value of $\lambda=\{1-\delta,1,1+\delta\}$. The weights are independently treated and the total energies of the systems are computed for each Hamiltonian  $H(\lambda)$. Of course, a total energy corresponding to a particular value of $\lambda$  should agree within statistical uncertainties with the result obtained by the standard DMC simulation using the Hamiltonian $H(\lambda)$.

In our calculations a single set of walkers is generated for all Hamiltonians $H(\lambda)$. The idea is to obtain the total energies $E^{\lambda}$ with correlated statistical fluctuations. The values obtained in this way allow us to perform the numerical computation of the derivatives involved in the application of the Feynman-Hellmann theorem with the required accuracy.

The drift of a given configuration $\vct{R}$ can be kept unique without any difficulty for all the values of $\lambda$. A new configuration $\vct{R}'$ is sampled from $\vct{R}$ according the standard procedure through

\begin{equation}
\label{eq:gd}
G_{d}\left(\vct{R},\vct{R}'\right)=\left(\frac{1}{4\pi D\Delta\tau}\right)^{-\frac{3N}{2}}\mbox{e}^{-\frac{1}{4D\Delta\tau}|\vct{R}-\vct{R}'-D\Delta\tau F\left(\vct{R}'\right)|^{2}},
\end{equation}

\noindent where $D=\frac{\hbar^{2}}{2m}$ is the diffusion constant, $\Delta \tau$ is the time step and $F=2\nabla\ln\Psi_{T}$.

In the calculation of the weights associated to a walker, required by the implementation of the Hellmann-Feynman theorem in the DMC method, a particular value of the parameter $E_{T}^{\lambda}$ is used for each value of $\lambda$. Weights are updated by the factor

\begin{equation}
G_{b}^{\lambda}\left(\vct{R},\vct{R}'\right)=\mbox{e}^{-\frac{\Delta\tau}{2}\left(E_{L}^{\lambda}\left(\vct{R}\right)+E_{L}^{\lambda}\left(\vct{R}'\right)-2E_{T}^{\lambda}\right)},
\end{equation}

\noindent when a configuration drifts from $\vct R$ to $\vct R'$. In this expression  $E_{L}^{\lambda}=H({\lambda})\Psi_{T}/\Psi_{T}$ are the local energies associated to the values of $\lambda$ we are considering. So, after the drift step, the weights are updated according to

\begin{equation}
\omega'^{\lambda}=\omega^{\lambda}G_{b}^{\lambda}\left(\vct{R},\vct{R}'\right).
\end{equation}

\noindent The parameters $E_{T}^{\lambda}$ are periodically changed so that the sum of the configurations weights for that particular $\lambda$ is kept approximately constant.

The energies at the $\ell$-th generation associated to each value of $\lambda$ are calculated in the usual way, as weighted averages over local energies of all the $i$ walkers

\begin{equation}
\label{eq:wa}
E^{\lambda}_{\ell}=\frac{\sum_{i}\omega^{\lambda}_{i}E^{\lambda}\left(\vct{R}_{i}\right)}{\sum_{i}\omega^{\lambda}_{i}}.
\end{equation}

\noindent Once the total energies are estimated, the kinetic and potential energies are computed through

\begin{equation}
V_{\ell} =\frac{E^{1+\delta}_{\ell}-E^{1-\delta}_{\ell}}{2\delta},
\qquad
K_{\ell}=E^{\lambda=1}_{\ell} - V_{\ell}.
\end{equation}

\noindent When the simulation reaches equilibrium these quantities are blocked and uncertainties estimated. In our calculations we have set $\delta=10^{-4}$.

\begin{table}[t]
\caption{Liquid phase results for the total $E_{T}$, kinetic $E_{K}$ and potential $E_{P}$ energies in units of Kelvin at the given density. The guiding function was of the Jastrow form.} \label{tab:Dados-DMC-Jastrow} 
\begin{indented}
\item[]
\begin{tabular}{@{}ccccccc}
\br
$\rho$ ($\sigma^{-3}$)  & ~~~~ & $E_{T}$ & ~~~~ & $E_{K}$ & ~~~~ & $E_{P}$ \\
\mr
0.3280 &  & -7.08$\pm$0.01 &  & 11.96$\pm$0.03 &  & -19.04$\pm$0.03 \\
0.3423 &  & -7.20$\pm$0.01 &  & 12.75$\pm$0.02 &  & -19.95$\pm$0.02 \\
0.3650 &  & -7.27$\pm$0.01 &  & 14.44$\pm$0.04 &  & -21.71$\pm$0.04 \\
0.3874 &  & -7.21$\pm$0.01 &  & 15.69$\pm$0.04 &  & -22.91$\pm$0.04 \\
0.4146 &  & -6.98$\pm$0.01 &  & 17.75$\pm$0.04 &  & -24.74$\pm$0.04 \\
0.4380 &  & -6.57$\pm$0.02 &  & 19.51$\pm$0.05 &  & -26.08$\pm$0.04 \\
\br
\end{tabular}
\end{indented}
\end{table}

So far all the additions we have performed to the DMC method are straightforward. Changes that require some elaboration concerns the combination of walkers, because those  regarding either splitting or keeping their weights are very easy. We adopt the following rules. $i)$ A walker is copied if all of its weights are greater than 2. Each copy bears half of the original  weights. This step can be repeated if needed. $ii)$ Walkers with at least one weight between a threshold $w_{thr}$ and 2 are kept without changes. $iii)$ Walkers $i$ and $j$ with all weights smaller than $w_{thr}$ are combined. For each weight $\lambda$ one of the usual rules of the literature is followed.  With probability $\frac{w_i^{\lambda}}{w_i^{\lambda}+w_j^{\lambda}}$ the weight $w_i^{\lambda}+w_j^{\lambda}$ is attribute to the walker $i$ and the value zero to the walker $j$. The sum of weights is attributed to the walker $j$  with probability $1 -  \frac{w_i^{\lambda}}{w_i^{\lambda}+w_j^{\lambda}}$ and the value zero to the walker $i$. For a particular $\lambda$ if the sum of the weights is equal to zero both walkers keep the value zero \cite{foot}. $iv)$ Delete a walker if all its weights are equal to zero.

The value of $w_{thr}$ is chosen by analyzing a compromise between two conflicting requirements. Ideally $w_{thr}$ needs to be small to avoid walkers with one or two of their weights equal to zero. These walkers are undesirable because they spoil the correlation we need to achieve for accurate numerical derivatives. On the other hand we do not want  to carry walkers with too low weights along the simulation. These kind of walkers would not contribute to the final result. In our simulations  we have verified that $w_{thr} = 0.3$ offers a good compromise between these two requirements.

It is worthwhile mentioning the procedure adopted in the trial energies update. Since we have to deal with three values of $E_{T}^{\lambda}$, it is useful to have an automatic scheme of updates. In this work the update was made every twenty generations using the following expression

\begin{equation}
\label{eq:et}
E_{T}^{\lambda}={E}^{\lambda}_b+\frac{C_{0}}{\tau_2 - \tau_1}\ln{\left(\frac{W^{\lambda}(\tau_2)}{W_{0}}\right)},
\end{equation}

\noindent where the first term $E^{\lambda}_b$ is the total energy averaged over a block of the last twenty generations, $C_0$ is a parameter which smooths the fluctuations of the weights sum in the calculation of the trial energies, $\tau_2 - \tau_1$ is the elapsed ``time'' interval since the last update, $W^{\lambda}$ is the sum of the $\omega^{\lambda}_i$ weights of all walkers at ``time'' $\tau_2$ and $W_{0}$ is another constant. It is a target value for the sum of  weights, the value we want to keep in the simulation at equilibrium. Its value is approximately equal to the number of walkers kept during the simulation. The update of the trial energies according to (\ref{eq:et}) was useful in the sense that we were able to perform runs where all the walkers did not have any of  their weights equal to zero. The expression of (\ref{eq:et}) is a reminiscent from the estimation of the energy by the growth estimator,

\begin{equation}
E_{g}^{\lambda}=E^{\lambda}_T+\frac{1}{\tau_2 - \tau_1 }\ln{\left(\frac{W^{\lambda}\left(\tau_1\right )}{W^{\lambda}\left(\tau_2 \right)}\right)},
\end{equation}

\noindent associated to the fluctuation of the  weights sum in the elapsed ``time'' interval $\tau_2 - \tau_1$. A ``time'' $\tau$ has a simple relation with the performed number of generations $M$ (attempts to move all particles) through $\tau_i = M_i\Delta\tau$, where $\Delta\tau$ is the time step used in (\ref{eq:gd}).

\section{Results}

\subsection{Liquid phase}

\begin{figure}[t]
\centering 
\includegraphics[scale=1]{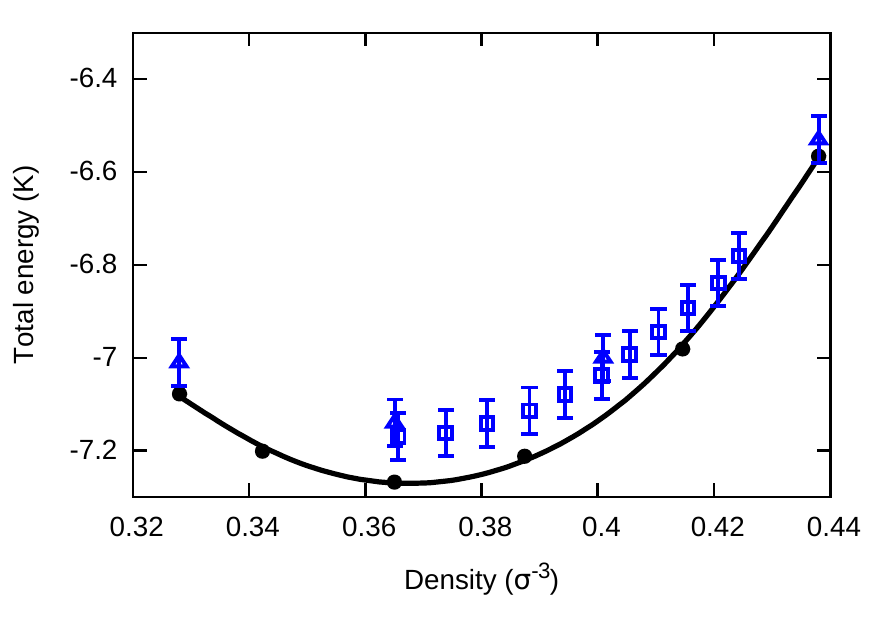}
\caption{\label{Fig:et_li} Equation of state as a function of the density in the liquid phase. Theoretical results are displayed by
\begingroup
\fontsize{12pt}{20pt}\selectfont $\bullet$
\endgroup
. The curve is a fit to the estimated values of the energies (see text). The errors are smaller than the symbol size. Experimental data for the liquid phase \cite{bru87,azi73} are shown by {\color{blue} $\boldsymbol{\square}$} and {\color{blue} $\boldsymbol{\triangle}$} symbols.}
\end{figure}

\begin{figure}[t]
\centering 
\includegraphics[scale=1]{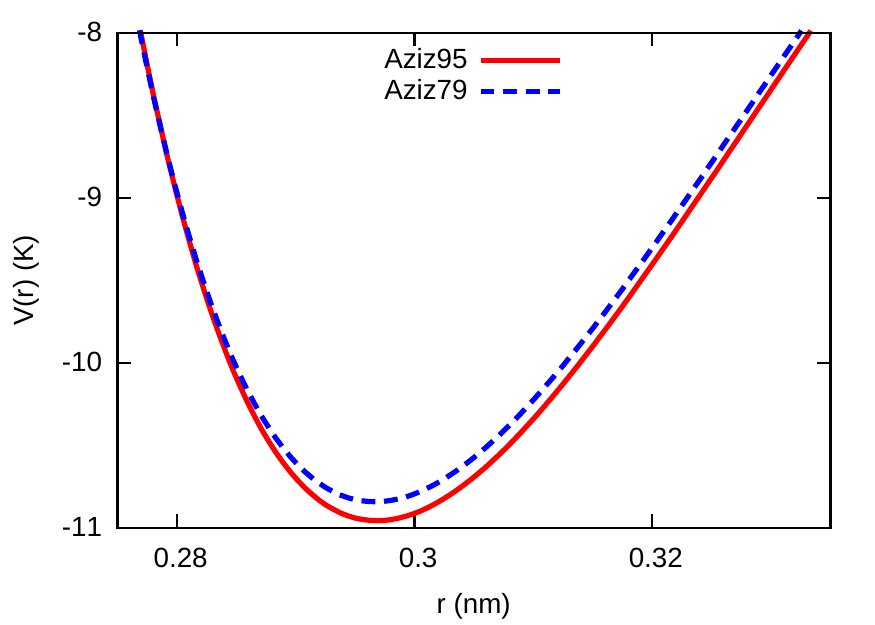}
\caption{\label{Fig:pot}
Comparison of the inter-atomic potential \textsc{hfd-b3-fci1} \cite{azi95} employed in this work (dubbed Aziz 95 at the Fig.\ legend) with the widely used \textsc{hfdhe2} \cite{azi79} (Aziz 79 at the Fig.\ key).}
\end{figure}

Estimates of the total energy per $^4$He atom are displayed in \tab{Dados-DMC-Jastrow}. These results, as a function of the density, are shown in \fig{et_li} as well. The curve in \fig{et_li} is a third degree polynomial fit to the estimated energies in the variable $(\rho - \rho_0)/\rho_0$ that depends on the fitted parameter $\rho_0$, the equilibrium density. Our fit gives $\rho_0 = 0.366 \pm 0.002$, in excellent agreement with experiment. In general the total energies are also in good agreement with experimental data obtained at finite temperatures. A possible bias in the results (that produces an underestimation of the experimental data) can be attributed to the lack of three-body interactions in the inter-atomic potential we have considered in our calculations \cite{uje07,uje06c}. In \fig{pot} we show a comparison between the inter-atomic potential used in this work and the one dubbed \textsc{hfdhe2} \cite{azi79} widely used in the literature. The former inter-atomic potential is more attractive than the last one and certainly a difference in energy can be attributed to the characteristics of these interactions.

\begin{figure}[t]
\centering  
\includegraphics{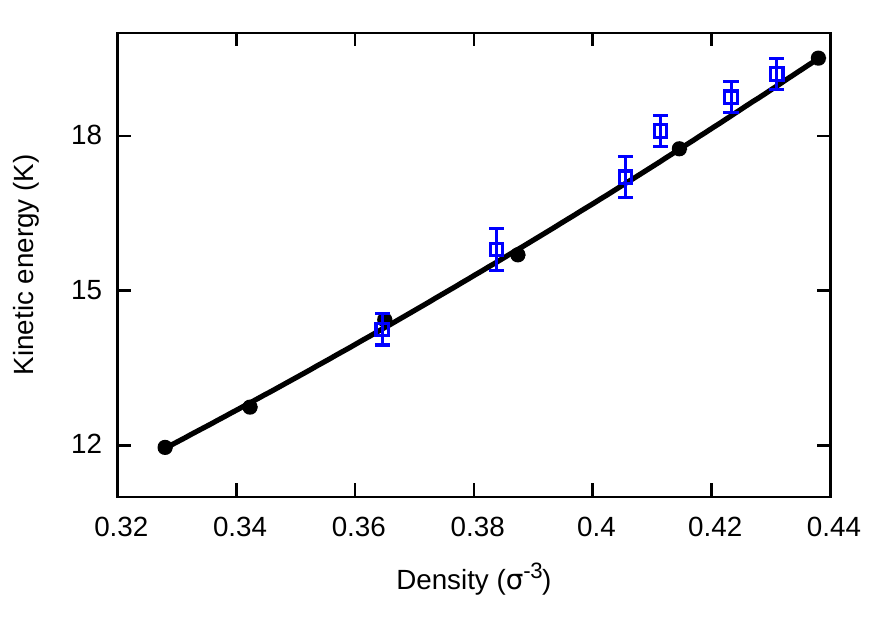} 
\caption{\label{Fig:ek_li} 
Kinetic energy per atom as a function of the density in the liquid phase. The curve stands for a fit to the estimated values displayed as
\begingroup
\fontsize{12pt}{20pt}\selectfont $\bullet$%
\endgroup
. Errors are smaller than the symbols size. The symbols {\color{blue} $\boldsymbol{\square}$} represent experimental results \cite{gly11}.}
\end{figure}

As we have discussed the total, potential and kinetic energies at a given density were estimated using correlated estimates of the desired quantities. In \fig{ek_li} we present the kinetic energy results of \tab{Dados-DMC-Jastrow}. The estimated values were fitted to a parabola. We considered a quadratic fit because this behavior was found in experimental data obtained at finite temperature \cite{baf95}. Even if we have in mind that this behavior was determined above the $\lambda$-transition. The experimental values displayed in the same figure were determined at $T$=0.05 K \cite{gly11}.

\subsection{Solid phase}

\begin{figure}[t]
\centering 
\includegraphics[scale=1]{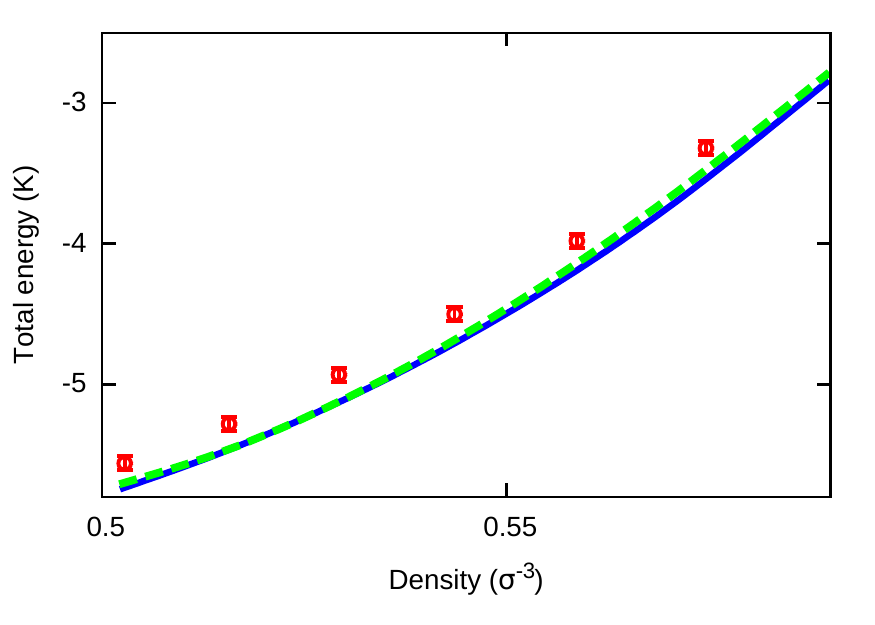}
\caption{\label{Fig:et_so}Total energy per atom as a function of the density in the solid phase. Theoretical results are displayed by curves fitted to the estimated values of the energies (see text). Results obtained with the NJ (blue line) and the SNJ (dashed and thick green line) guiding functions are hard to distinguish at the figure resolution. Experimental data \cite{edw65,ada07,bla93} are displayed by
\begingroup
{\color{red} \fontsize{12pt}{20pt}\selectfont $\boldsymbol{\circ}$}%
\endgroup
, their errors have been assumed to be half of the penultimate less significative figure.}
\end{figure}

A defect free \textsl{hcp} crystalline structure with $N=288$ $^4$He atoms simulation cell was used to estimate the total energy per atom in the solid phase. In \fig{et_so} the results (as a function of the density) are shown by curves fitted to the values of the total energies displayed in \tab{Dados-DMC-1-1}. In this phase, at the level of precision we have considered, the results obtained with a guiding function of the Nosanov-Jastrow form are indistinguishable within statistical uncertainty or are marginally lower than those determined with the symmetric Nosanov-Jastrow guiding function. The only exception is at $\rho=0.5277 \sigma^{-3}$ where the SNJ guiding function gives a marginally lower energy. The overall situation might be attributed to the different degrees of superposition that the guiding functions have with respect to the true ground state of the system. The equation of state is also displayed by curves that are third degree polynomial fits to the estimated energies in the variable $(\rho - \rho_0)/\rho_0$ that depends on the fitted parameter $\rho_0$. The parameter $\rho_0$ does not have any particular meaning in this phase. Once again the total energies are in general in good agreement with the experimental data obtained at finite temperatures. The same bias observed in the liquid phase can be seen in the solid phase, and here also it is attributed to the lack of three-body interactions in the inter-atomic potential considered \cite{uje07,uje06c}.

\begin{table}[t]
\caption{Total $E_{T}$, kinetic $E_{K}$ and potential $E_{P}$ energies in units of Kelvin for the solid phase at the given densities. The results were obtained with the Nosanov-Jastrow (NJ) and the symmetric Nosanov-Jastrow (SNJ) guiding functions.} \label{tab:Dados-DMC-1-1}
\begin{indented}
\item[]
\begin{tabular}{@{}ccccccc}
\br
$\rho$ ($\sigma{}^{-3}$) & \multicolumn{2}{c}{$E_{T}$} & \multicolumn{2}{c}{$E_{K}$} & \multicolumn{2}{c}{$E_{P}$} \\
\mr
 & \multicolumn{1}{c}{NJ} & \multicolumn{1}{c}{SNJ} & \multicolumn{1}{c}{NJ} & \multicolumn{1}{c}{SNJ} & \multicolumn{1}{c}{NJ} & \multicolumn{1}{c}{SNJ} \\
0.5026 & -5.74$\pm$0.01 & -5.70$\pm$0.01 & 26.37$\pm$0.04 & 26.61$\pm$0.05 & -32.11$\pm$0.04 & -32.31$\pm$0.05 \\
0.5126 & -5.54$\pm$0.01 & -5.53$\pm$0.02 & 27.28$\pm$0.04 & 27.40$\pm$0.10 & -32.82$\pm$0.03 & -33.00$\pm$0.10 \\
0.5277 & -5.16$\pm$0.01 & -5.20$\pm$0.01 & 28.66$\pm$0.06 & 28.73$\pm$0.05 & -33.83$\pm$0.06 & -33.93$\pm$0.04 \\
0.5344 & -4.98$\pm$0.01 & -4.93$\pm$0.01 & 29.37$\pm$0.03 & 29.85$\pm$0.05 & -34.36$\pm$0.03 & -34.78$\pm$0.05 \\
0.5494 & -4.51$\pm$0.01 & -4.50$\pm$0.02 & 30.76$\pm$0.07 & 30.99$\pm$0.07 & -35.28$\pm$0.08 & -35.49$\pm$0.06 \\
0.5694 & -3.78$\pm$0.03 & -3.70$\pm$0.01 & 32.80$\pm$0.10 & 33.15$\pm$0.05 & -36.60$\pm$0.10 & -36.86$\pm$0.05 \\
0.5895 & -2.85$\pm$0.01 & -2.80$\pm$0.01 & 34.81$\pm$0.02 & 35.10$\pm$0.10 & -37.67$\pm$0.02 & -37.90$\pm$0.10 \\
\br
\end{tabular}
\end{indented}
\end{table}

\begin{figure}[t]
\centering  
\includegraphics{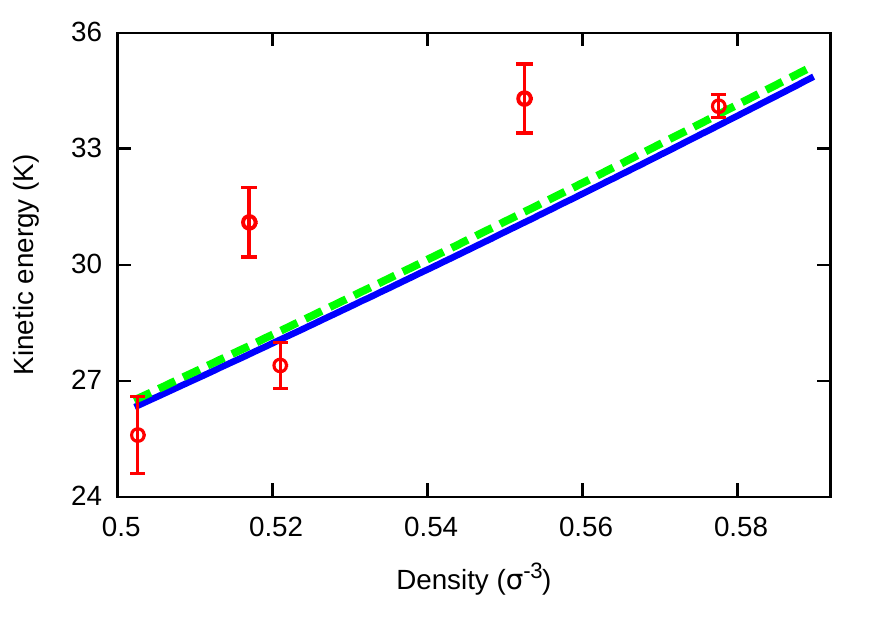} 
\caption{\label{Fig:ek_so} Kinetic energy per atom as a function of the density in the solid phase. The curves stand for fits to the estimated values (see text). Experimental data \cite{bla93,ada07,dia04,hil84} are displayed by
\begingroup
{\color{red} \fontsize{12pt}{20pt}\selectfont $\boldsymbol{\circ}$}%
\endgroup
. Fits to the estimated values obtained with the NJ (blue line) and the SNJ (dashed and thicker green line) guiding functions almost coincide.}
\end{figure}

Correlated estimates of the quantities needed to estimate the kinetic energy were used and the values obtained are presented in \fig{ek_so}. In this phase a straight line was used to fit the kinetic energy results of \tab{Dados-DMC-1-1}. We considered a linear fit since this behavior was found in experimental data obtained at temperatures above the $\lambda$-transition \cite{baf95}. The experimental values displayed in the figure are from \cite{edw65,ada07,bla93}. Although in the solid phase the kinetic energies obtained with a Nosanov-Jastrow guiding function could be indistinguishable (within statistical uncertainty) from those of a symmetric Nosanov-Jastrow at a few densities, always the results from the first guiding function were lower than those obtained with the second one. This trend is not obvious, since atoms are in principle much less localized when the simulation is guided by the symmetric Nosanov-Jastrow guiding function.

\section{Final Comments}

The multi-weight DMC method has shown to be reliable to compute the kinetic energies of systems formed by helium atoms. Furthermore it is not hard to realize that there are other applications of the multi-weight DMC method. One of them in qhich we are working is its application in the refinement of  guiding functions of systems that obey Fermi statistics without the need of variational approaches. This would be possible if small changes in a guiding function could be mapped into changes of the inter-atomic potential. We already know that the multi-weight method can be used to analyze the consequences of small changes in the interacting potential \cite{uje07,uje06}.

In the liquid phase as expected the agreement with experiment is very good. The kinetic energy in this phase at all densities we have considered are also in good agreement with experiment and other theoretical estimates of this quantity.

Our results in the solid phase show that both guiding functions, the Nosanov-Jastrow and the symmetric version \cite{caz09} we have considered, can give energies in good agreement with experiment. However, the trend we have observed for the estimated values of the energies suggests that the Nosanov-Jastrow guide function can be the preferred form when a simple DMC calculation needs to be performed. Although the differences in the results obtained with these guiding functions might be attributed to the particular and well know fact that exchange is not so important in solid helium, the result could also be a consequence of the approximation used in the symmetric Nosanov-Jastrow function. It does not have all the terms that a true symmetrization would give.

In conclusion, the extension of the DMC method we have applied in this work allows an accurate determination of quantities that depends on differences of estimates even when these differences are very small. The method avoids the use of quantities with uncorrelated statistical uncertainties to compute  their differences. This is a well known difficulty of Monte Carlo methods that can lead to a complete loss of accuracy of the quantities of interest. Moreover and more important, the application of this method in slightly different contexts may bring an enhancement of our understanding of the quantum many-body systems.

\ack The authors acknowledge financial support from the Brazilian agencies \textsc{fapesp}, \textsc{faperj} and \textsc{cnp}q. Part of the computations were performed at the \textsc{cenapad} high-performance computing facility at Universidade Estadual de Campinas.


\begin{thebibliography}{99}
\bibitem{foot}
Just before completion of this work, we have realized that walkers could be combined in a slightly improved manner. Our results can have less fluctuations if we chose to combine a walker with all of its weights smaller than $w_{tresh}$ with another one that has all of them greater than 2
\end{thebibliography}
\end{document}